\documentclass[12pt]{article}
\usepackage{amsfonts,amssymb}
\usepackage{stmaryrd}
\begin{document}

\title{Relation of deformed nonlinear algebras with linear ones }

\author{
A. Nowicki$^1$\footnote{A.Nowicki@if.uz.zgora.pl}, V. M. Tkachuk$^2$\footnote{tkachuk@ktf.franko.lviv.ua}\\
$^1$Institute of Physics, University of Zielona G\'ora \\
Prof. Z. Szafrana 4a, 65-516 Zielona G\'ora, Poland \\
$^2$Department of Theoretical Physics,\\ Ivan Franko National
University of Lviv,\\ 12 Drahomanov St., Lviv, UA-79005, Ukraine}

\maketitle

\begin{abstract}
The relation between nonlinear algebras and linear ones
is established. For one-dimensional nonlinear deformed Heisenberg algebra with two operators we
find the function of deformation for which this nonlinear algebra can be transformed to a linear one
with three operators. We also establish the relation between Lie algebra of total angular momentum
and corresponding nonlinear one. This relation gives a possibility to simplify and to solve
the eigenvalue problem for the Hamiltonian in a nonlinear case using the reduction of this problem to the case
of linear algebra. It is demonstrated on the example of harmonic oscillator.
\end{abstract}

\section{Introduction}

Historically the first publication on the subject of deformed Heisenberg algebra was Snyder's paper \cite{Snyder47} where
the Lorentz-covariant
deformed Heisenberg algebra leading to a quantized
space-time was proposed. For a long time there were only a few papers on this
subject. The interest to deformed algebras was renewed after
investigations in string theory and quantum gravity which suggested the
existence of a nonzero minimal uncertainty in position following
from the generalized uncertainty principle (GUP) (see, e.g., \cite{gross, maggiore, witten}).
In \cite{maggiore2,Kem95,Kem96} it was shown that GUP and nonzero minimal
uncertainty in position can be obtained from some modifications of Heisenberg
algebra. Subsequently there were published many
papers where different systems and their properties in space with deformed
Heisenberg algebra was studied: one-dimensional harmonic
oscillator with minimal uncertainty in position \cite{Kem95} and
also with minimal uncertainty in position and momentum
\cite{Tkachuk1,Tkachuk2}, $D$-dimensional isotropic harmonic
oscillator \cite{chang, Dadic}, three-dimensional Dirac oscillator
\cite{quesne} and one-dimensional Coulomb problem \cite{fityo},
(1+1)-dimensional Dirac oscillator with Lorentz-covariant deformed
algebra \cite{Quesne10909}, three-dimensional Coulomb problem with
deformed Heisenberg algebra in the frame of perturbation theory
\cite{Brau,Benczik,mykola,Stet,mykolaOrb}, singular inverse square
potential with a minimal length \cite{Bou1,Bou2},
the scattering problem in the deformed space with minimal length \cite{Stet07},
ultra-cold
neutrons in gravitational field with minimal length
\cite{Bra06,Noz10,Ped11},
the influence of minimal length on Lamb's shift, Landau levels, and tunneling current in scanning tunneling microscope \cite{Das,Ali2011},
the Casimir effect in a space with minimal length \cite{Frassino},
the effect of noncommutativity and of the existence of a minimal length on the phase space of cosmological model \cite{Vaki}.
various physical consequences which follow from the noncommutative Snyder space-time geometry \cite{Batt},
the classical mechanics in a space with deformed Poisson brackets \cite{BenczikCl,Fryd,Sil09},
composite system in deformed space with
minimal length \cite{Quesne10,Bui10}, equivalence principle in deformed space with
minimal length \cite{Tka12}.

We would like to note that all the deformed algebras studied in the above references are nonlinear ones.
In this paper we consider the relation between nonlinear algebras and linear ones. As far as we know, up to now
this question
has not been studied in the literature. So, the
aim of this paper is to fill this gap.
\section{Linearization of deformed algebra}
Let us consider a one-dimensional deformed nonlinear Heisenberg algebra with function of deformation $f(P)$,
namely
\begin{eqnarray} \label{a1}
[X,P]=if(P),
\end{eqnarray}
where $\hbar=1$.
The function of deformation is an even  $f(-P)=f(P)$ and positive function. It means that the space has the same properties in two opposite
directions.

The momentum representation reads:
\begin{eqnarray} \label{repr0}
P=p, \ \ X=if(p){d\over dp}
\end{eqnarray}
and acts on square integrable functions $\phi(p) \in \mathcal{L}^2(-a\,,a;f)\, ,(a
\leq \infty)$ where the norm of $\phi$ is given by
\begin{equation}
\parallel \phi\parallel^2 \ = \ \int_{-a}^a\,\frac{dp}{f(p)} \mid\phi(p)\mid^2\, .\label{aa3}
\end{equation}
For the hermiticity of ${X}$ it is enough that $\phi(-a) = \phi(a)$ or $\phi(-a) = -\phi(a)$. More strong boundary conditions
$\phi(-a) = \phi(a)=0$ were considered in \cite{Mas12}.

Now we extend this algebra by one additional operator $F=f(p)$. Using representation (\ref{repr0}) one can easily find
\begin{eqnarray}
[X,F]=[if{d\over dp},f(p)]=iff',\ \ [P,F]=[p,f(p)]=0.
\end{eqnarray}
We require that algebra of three operators $X$, $P$ and $F$ is linear and to close this algebra we put
\begin{eqnarray}\label{eqf}
ff'=\alpha +\beta p + \gamma f.
\end{eqnarray}
where $\alpha$, $\beta$ and $\gamma$ are real parameters.
Note that the linear combination in the right hand side of (\ref{eqf}) does not contain $X$ because $ff'$ is a function of $p$ only.

Let us take into account the fact that the function of deformation is an even one. Then changing $p\to-p$ in (\ref{eqf})
and taking into account that  $f(-p)=f(p)$ we find
\begin{eqnarray}\label{eqfN}
ff'=-\alpha +\beta p - \gamma f.
\end{eqnarray}
Comparing (\ref{eqf}) and (\ref{eqfN}) we find $\alpha=\gamma=0$. So, only equation
\begin{eqnarray}\label{eqfNN}
ff'=\beta p.
\end{eqnarray}
has even solutions.

The solution in this case reads
\begin{eqnarray}\label{solf1}
f(p)=\pm\sqrt{c+\beta p^2},
\end{eqnarray}
where $c$ is the constant of integration.  Choosing
$"+"$ in (\ref{solf1}) and putting
constant of integration $c=1$ we have an even and positive function of deformation
\begin{eqnarray}\label{solf11}
f(p)=\sqrt{1+\beta p^2}.
\end{eqnarray}
 The linear algebra in this case reads
\begin{eqnarray} \label{lin-a1}
[X,P]=iF, \ \ [X,F]=i\beta P, \ \ [P,F]=0.
\end{eqnarray}
This is Lie algebra. One can find the Casimir operator (invariant) for this algebra
\begin{eqnarray}\label{Kaz1}
K=P^2-{1\over\beta}F^2
\end{eqnarray}
commuting with all elements of the algebra. When we return to nonlinear deformed algebra we find that
the Casimir operator is constant. Really,
\begin{eqnarray}
K=p^2-{1\over\beta}f^2(p)=-{1\over\beta}.
\end{eqnarray}

So, in this section we found that when the function of deformation is given by (\ref{solf11}) then
nonlinear algebra (\ref{a1}) for two operators can be transformed to linear algebra (\ref{lin-a1}) with three operators.

Let us consider in more detail linear algebra (\ref{lin-a1}) denoting the hermitean generators as
\begin{equation}
A_1 \ = \ \lambda\,P\, ,\quad A_2 \ = \ F\, ,\quad A_3 \ = \ \frac{1}{\lambda}\,X\, ,\label{r1.1}
\end{equation}
and they fulfill commutation relations in the form
\begin{equation}
\left[ A_1\,,A_2\right] \ = \ 0\, ,\quad \left[ A_3\,,A_1\right] \ = \ i\,A_2\, ,\quad \left[ A_3\,, A_2\right] \ = \  i\,\mbox{sign}(\beta)\,A_1\, ,\label{r1.2}
\end{equation}
where $\beta = \pm \lambda^2$. So, the momentum realization of this algebra  takes a canonical form
\begin{equation}
A_1 \ = \ \lambda\,p\, ,\quad A_2 \ = \ \sqrt{1 + \beta\,p^2}\, ,\quad A_3 \ = \ \frac{i}{\lambda}\,\sqrt{1 + \beta\,p^2} \frac{d}{dp}\, .\label{r1.3}
\end{equation}
It is convenient to use a pair of commuting hermitean operators $P_+\,,P_-$ defined as follows
\begin{eqnarray}
P_+ &=& A_1 + A_2 \ = \  \lambda\,P + F\, ,\label{r1.201}\\ P_- &=&   A_2 - A_1 \ = \ F - \lambda\,P \, ,\label{r1.4}
\end{eqnarray}
satisfying the relations
\begin{eqnarray}
\left[ A_3\,, P_+\right] &=& i\,\mbox{sign}(\beta) \left( A_1 + \mbox{sign}(\beta)\,A_2\right)\, ,\label{r1.5}\\
\left[ A_3\,, P_-\right] &=& i\,\mbox{sign}(\beta) \left( A_1 - \mbox{sign}(\beta)\,A_2\right)\, .\label{r1.6}
\end{eqnarray}
In this basis the Casimir operator (\ref{Kaz1}) take the form
\begin{equation}
K(\beta) \ = \ P^2 - \frac{1}{\beta}\,F^2=\left\{\begin{array}{l} C_2(\lambda^2) \equiv\frac{1}{\lambda^2}\,P_+ P_-\,;\quad \beta=\lambda^2 > 0,\\ \\ C_2(-\lambda^2) \equiv\frac{1}{2 \lambda^2}\left( P_+^2 + P_-^2\right)\,;\beta= -\lambda^2 < 0\,.\end{array}\right.\label{r1a.1}
\end{equation}

\noindent(i)\quad{\it Inhomogeneous hyperbolic rotations of two-dimesional plane $ISO(1\,,1)$}\medskip

In the case $\beta > 0$ we obtain three dimensional Lie algebra generated by $P_+\,,P_-\,,A_3$ satisfying the commutation relations
\begin{equation}
\left[ A_3\,, P_+\right] \ = \ i\,P_+\, ,\quad \left[ A_3\,, P_-\right] \ = \ -i\,P_-\, ,\quad \left[P_+\,,P_-\right] \ = \ 0\, ,\label{r1.7}
\end{equation}
where $A_3$ generates $\mathfrak{so}(1\,,1)$ hyperbolic rotations. The Lie algebra $\mathfrak{iso}(1,1) \sim \mathfrak{so}(1,1)\supsetplus \mathfrak{t}^2$ is a semidirect sum of hyperbolic rotations and translations $\mathfrak{t}^2$ (we use the notation of reference \cite{Gilmore}).

We can change the momentum variable $\lambda\,p = \sinh(\lambda\xi)$ then generators $P_\pm$ take a simple form
\begin{eqnarray}
P_+ &=& \sqrt{1 + \lambda^2\,p^2} + \lambda\,p \ = \ e^{\lambda\,\xi} \ = \ e^{\mbox{\scriptsize{arcsh}}(\lambda P)}\, ,\label{r1.8}\\
P_- &=& \sqrt{1 + \lambda^2\,p^2} - \lambda\,p \ = \ e^{-\lambda\,\xi} \ = \ e^{-\mbox{\scriptsize{arcsh}}(\lambda P)}\, .\label{r1.9}
\end{eqnarray}
In this realization the spectrum of both generators $P_\pm$ is positive and  it is easy to find the hermitean generators of the algebra $\mathfrak{iso}(1,1)$ in the Hilbert space of the square integrable functions $\phi\in \mathcal{L}^2(\mathcal{R})$. First we notice that the scalar products are related as follows (cf.(\ref{aa3}))
\begin{equation}
\langle\psi\,,\phi \rangle \ = \ \int_{-\infty}^\infty\,\frac{dp}{\sqrt{1 + \lambda^2 p^2}}\,{\psi}^*(p)\,\phi(p) \ = \
\int_{-\infty}^\infty\,d\xi\,\tilde{\psi}^*(\xi)\,\tilde{\phi}(\xi) \, .\label{r1.10}
\end{equation}
where $\tilde{\phi}(\xi) = \phi((1/\lambda)\,\sinh \xi)$ and the generators of Lie algebra $\mathfrak{iso}(1,1)$ take the form
\begin{equation}
A_3 \ = \ \frac{i}{\lambda}\,\frac{d}{d\xi}\, ,\quad P_+ \ = \ e^{\lambda\,\xi}\, ,\quad P_- \ = \ e^{-\lambda\,\xi}\, .\label{r1.11}
\end{equation}
$$
$$
\noindent(ii)\quad{\it Inhomogeneous  rotations of two-dimesional plane $ISO(2)$}\medskip

Similarly, in the case $\beta < 0$ we get the Lie algebra $\mathfrak{iso}(2)$ of transformations of the Euclidean plane
\begin{equation}
\left[ A_3\,, P_+\right] \ = \ i\,P_-\, ,\quad \left[ A_3\,, P_-\right] \ = \ -i\,P_+\, ,\quad \left[P_+\,,P_-\right] \ = \ 0\, ,\label{r1.12}
\end{equation}
where $A_3$ is a generator of rotations of two-dimensional plane. The algebra $\mathfrak{iso}(2)\sim \mathfrak{so}(2)\supsetplus \mathfrak{t}^2$   is a semidirect sum of $\mathfrak{so}(2)$ rotations and two-dimensional translations $\mathfrak{t}^2$.

Introducing new variable $\theta\,, \lambda\,p = \sin(\lambda\,\theta)$ we get
\begin{eqnarray}
P_+ &=& \sqrt{1 - \lambda^2\,p^2} + \lambda\,p \ = \ \cos(\lambda\,\theta) + \sin(\lambda\,\theta)\, ,\label{r1.13}\\
P_- &=& \sqrt{1 - \lambda^2\,p^2} - \lambda\,p \ = \ \cos(\lambda\,\theta) - \sin(\lambda\,\theta)\, .\label{r1.14}
\end{eqnarray}
The hermicity condition puts on the generators $P_\pm$ implies that $-1 \leq \lambda\,p \leq 1$ and $-\frac{\pi}{2\lambda} \leq \theta \leq \frac{\pi}{2\lambda}$ if we demand that the correspondence $p \leftrightarrow \theta$ is unambiguous.
In considered case the scalar products are related in the following way
\begin{equation}
\langle\psi\,,\phi \rangle \ = \ \int_{-1/\lambda}^{1/\lambda}\,\frac{dp}{\sqrt{1 - \lambda^2 p^2}}\,{\psi}^*(p)\,\phi(p) \ = \
\int_{-\pi/2\lambda}^{\pi/2\lambda}\,d\theta\,\tilde{\psi}^*(\theta)\,\tilde{\phi}(\theta) \, ,\label{r1.15}
\end{equation}
where $\tilde{\phi}(\theta) = \phi((1/\lambda)\,\sin \theta)$ and the generators of Lie algebra $\mathfrak{iso}(2)$ of Euclidean group are given by the formulae
\begin{equation}
A_3 \ = \ \frac{i}{\lambda} \frac{d}{d \theta}\, ,\quad P_\pm \ = \ \cos(\lambda \theta) \pm \sin(\lambda \theta)\, ,\label{r1.16}
\end{equation}
or equivalently
\begin{equation}
A_1 \ = \ \frac{1}{2} \left(P_+ - P_-\right) \ = \ \sin(\lambda\,\theta)\, ,\quad A_2 \ = \ \frac{1}{2} \left(P_+ + P_-\right) \ = \ \cos(\lambda\,\theta)\, .\label{r1.17}
\end{equation}
It is worthwhile to notice that operators $A_1\,, A_2$ in this representation appeared in quantum mechanics as the sine and cosine operators in discussion on existence of the self-adjoint phase operator \cite{Nieto}.\medskip

\noindent(iii)\quad{\it On expansion to simple algebras $\mathfrak{so}(2, 1)$ and $\mathfrak{so}(3)$.}\medskip

Both considered Lie algebras are closely related to three dimensional orthogonal and pseudo-orthogonal rotation algebras  by the expansion procedure.
Further we shall follow the framework of expansion described in ref.\cite{Gilmore1} (Chapter 10).

We define new generators as follows
\begin{equation}
\Pi_\pm \ = \ \left[ A_3^2\,, P_\pm\right] \ = \ \left\{\begin{array}{l} \pm i \{A_3\,, P_\pm\}\quad\mbox{for}\,;\quad \beta=\lambda^2 > 0,

\\ \\ \pm i \{A_3\,, P_\mp\}\quad\mbox{for}\quad\beta= -\lambda^2 < 0\,.\end{array}\right.\label{r1a.2}
\end{equation}
where $\{A\,,B\} = A B + B A$. Using the relation $\{A\,,B\}= 2AB - \left[ A\,,B\right]$ we find the commutation relations\\
(a)\quad for $\beta = \lambda^2 > 0$:
\begin{eqnarray}
\left[ A_3\,, \Pi_\pm \right] &=& - \{ A_3\,, P_\pm \} \ = \ \pm i \Pi_\pm\, ,\label{r1a.3}
\\
\left[ \Pi_+\,, \Pi_- \right] &=& -8i\,A_3 P_+ P_- \ = \ -8i\lambda^2\,A_3 C_2(\lambda^2)\, .\label{r1a.4}
\end{eqnarray}
Introducing redefined generators
\begin{equation}
\tilde{P}_\pm(\varepsilon) \ = \ \frac{\Pi_\pm}{2\sqrt{\varepsilon P_+ P_-}} \ = \ \frac{\Pi_\pm}{2\sqrt{\varepsilon\lambda^2 C_2(\lambda^2)}}\, ,\label{r1a.5}
\end{equation}
where $\varepsilon = \pm 1$ and the Casimir is given by (\ref{r1a.1}), then we obtain the relations
\begin{equation}
\left[ A_3\,, \tilde{P}_\pm(\varepsilon) \right] \ = \ \pm i\,\tilde{P}_\pm(\varepsilon)\, ,\qquad \left[\tilde{P}_+(\varepsilon)\,, \tilde{P}_-(\varepsilon) \right] \ = \ -2i\varepsilon\,A_3\, ,\label{r1a.6}
\end{equation}
or
\begin{equation}
\left[ \tilde{A}_1\,, \tilde{A}_2\right] \ = \ -i \varepsilon\,\tilde{A}_3\, ,\quad
\left[ \tilde{A}_2\,, \tilde{A}_3\right] \ = \ i\,\tilde{A}_1\, ,\quad
\left[ \tilde{A}_3\,, \tilde{A}_1\right] \ = \ i\,\tilde{A}_2\, ,\label{r1a.7}
\end{equation}
where $\tilde{P}_\pm(\varepsilon) = \tilde{A}_2 \pm \tilde{A}_1\, , \tilde{A}_3 = A_3$. Let us notice that for $\varepsilon = 1$ all generators $\tilde{A}_k ( k=1,2,3)$ are hermitean and the second order Casimir operator have the form
\begin{equation}
\tilde{C}_2(\varepsilon) \ = \ \tilde{A}_1^2 + \tilde{A}_2^2 - \varepsilon \tilde{A}_3^2\, ,\label{r1a.8}
\end{equation}
and for $\varepsilon=1$ the hermitean generators $\tilde{A}_k$  span the Lie algebra $\mathfrak{so}(2,1)$. On the other hand for $\varepsilon=-1$ the generators $\tilde{A}_1\,,\tilde{A}_2$ become antihermitean operators and span the Lie algebra $\mathfrak{so}(3)$ as an extension of $\mathfrak{so}(2,1)$ algebra by the Weyl unitary trick \cite{Gilmore1}.\medskip\\Similarly, the commutation relations take the form\\(b) for $\beta = -\lambda^2 < 0$:
\begin{eqnarray}
\left[ A_3\,, \Pi_\pm \right] &=& \pm i\,\Pi_\mp\, ,\label{r1b.3}\\
\left[ \Pi_+\,, \Pi_- \right] &=& 4i\,A_3 \left( P_+^2 + P_-^2\right) \ = \ 8\lambda^2 i\,A_3 C_2(-\lambda^2)\, .\label{r1a.9}
\end{eqnarray}
Now, redefined generators are given by
\begin{equation}
\tilde{P}_\pm(\varepsilon) \ = \ \frac{\Pi_\pm}{\sqrt{2\varepsilon\left( P_+^2 + P_-^2\right)}} \ = \ \frac{\Pi_\pm}{2\sqrt{\varepsilon\lambda^2 C_2(-\lambda^2)}} \, ,\label{r1a.10}
\end{equation}
satisfying the relations
\begin{equation}
\left[ A_3\,, \tilde{P}_\pm(\varepsilon) \right] \ = \ \pm i\,\tilde{P}_\mp(\varepsilon)\, ,\qquad \left[\tilde{P}_+(\varepsilon)\,, \tilde{P}_-(\varepsilon) \right] \ = \ 2i\varepsilon\,A_3\, ,\label{r1a.11}
\end{equation}
The commutation relations for $\tilde{A}_k$ generators take the form
\begin{equation}
\left[ \tilde{A}_1\,, \tilde{A}_2\right] \ = \ i \varepsilon\,\tilde{A}_3\, ,\quad
\left[ \tilde{A}_2\,, \tilde{A}_3\right] \ = \ i\,\tilde{A}_1\, ,\quad
\left[ \tilde{A}_3\,, \tilde{A}_1\right] \ = \ i\,\tilde{A}_2\, ,\label{r1a.12}
\end{equation}
and the Casimir operator is given by
\begin{equation}
\tilde{C}_2(\varepsilon) \ = \ \tilde{A}_1^2 + \tilde{A}_2^2 + \varepsilon \tilde{A}_3^2\, .\label{r1a.13}
\end{equation}
We see that for $\varepsilon=1$ the generators $\tilde{A}_k$ are hermitean operators and form a simple three dimensional rotation algebra  $\mathfrak{so}(3)$. For $\varepsilon <0$ the generators $\tilde{A}_1, \tilde{A}_2$ become nonhermitean and we obtain a nonhermitean realization of $\mathfrak{so}(2,1)$ non-compact pseudo-orthogonal Lie algebra.

\subsection{Minimal length versus discreteness of configurational space}

In the paper \cite{Mas12}
we obtain the result for the minimal length which is defined as the minimum in uncertainty of position
operator
\begin{eqnarray}
l_0={\rm min}\sqrt{\langle(\Delta X)^2\rangle}
= \ \frac{\pi}{2}\,\left(\int_0^{a}{d p \over
f(p)}\right)^{-1}\, .
\end{eqnarray}
With function of deformation (\ref{solf11})
for $\beta\ge 0$ the momentum is defined on the whole line
$-\infty<p<\infty$ and we demand the boundary conditions $\phi(-\infty)=\phi(\infty)=0$.
The minimal length $l_0$ in this case
is zero.

Let us consider in more detail the second case $\beta=-\lambda^2<0$ with the function of deformation
\begin{eqnarray}\label{solf12}
f(p)=\sqrt{1-\lambda^2 p^2}.
\end{eqnarray}
The momentum now is bounded to the region $-1/\lambda< p<1/\lambda$.
In this case the minimal length $l_0$ is nonzero (see paper \cite{Mas12})
\begin{eqnarray}\label{minl1}
\sqrt{\langle(\Delta X)^2\rangle}\ge l_0=\lambda.
\end{eqnarray}
Here it is important to note that the result concerning the minimal length
was obtained in \cite{Mas12} for zero boundary conditions
\begin{eqnarray}\label{bc0}
\phi(-1/\lambda) = \phi(1/\lambda)=0.
\end{eqnarray}
The eigenvalue equation for the position operator $X$ with these boundary conditions does not have solutions
that are in agreement with the the fact that according to (\ref{minl1}) the uncertainty in position is nonzero $\langle(\Delta X)^2\rangle\ne0$.

Now let us consider nonzero boundary conditions
\begin{eqnarray}\label{bcn0}
\phi(-1/\lambda) = \phi(1/\lambda)
\end{eqnarray}
and
\begin{eqnarray}\label{bcn0m}
\phi(-1/\lambda) = -\phi(1/\lambda)
\end{eqnarray}
which are weaker than those imposed in \cite{Mas12}. Therefore the result concerning the minimal length now is not applied.

The eigenvalue equation for the position operator in this case reads
\begin{eqnarray}
i\sqrt{1-\lambda^2p^2}{d\over dp}\phi(p)=l\phi(p).
\end{eqnarray}
The square integrable solution for eigenfunction is
\begin{eqnarray}\label{eisol1}
\phi(p)=\sqrt{\lambda\over\pi}\exp\left(-i{l\over\lambda}\arcsin\lambda p\right).
\end{eqnarray}
The boundary condition (\ref{bcn0}) in an explicit form reads
\begin{eqnarray}
\exp\left(-i{l\over\lambda}\pi\right)=1
\end{eqnarray}
and for the eigenvalues we find
\begin{eqnarray}\label{eigval1p}
l_n=\lambda 2n, \ \ n=0,\pm 1, \pm 2, ...
\end{eqnarray}
Similarly, for boundary condition (\ref{bcn0m}) we find
\begin{eqnarray}
\exp\left(-i{l\over\lambda}\pi\right)=-1
\end{eqnarray}
and the eigenvalues now read
\begin{eqnarray}\label{eigval1m}
l_n=\lambda (2n+1), \ \ n=0,\pm 1, \pm 2, ...
\end{eqnarray}

So, the eigenvalues of position operator are discrete.
It is obvious that for eigenstates (\ref{eisol1}) the uncertainty in position is zero and therefore the minimal length is zero.

Thus we have two possible scenarios. When boundary condition (\ref{bc0})
is imposed then there exists nonzero minimal length and there are no solutions of the eigenvalue equation for the position operator.
For boundary condition (\ref{bcn0}) and (\ref{bcn0m}) the solutions of the
eigenvalue equation for the position operator exist, the eigenvalues are discrete and the minimal length is zero.
\subsection{Angular momentum representation of linear algebra}
It is easy to verify that linear algebra (\ref{lin-a1}) with $\beta=-\lambda^2$ is satisfied by the operators
\begin{eqnarray}\label{XPF}
X=\lambda\left(-ix{\partial\over\partial y}+iy{\partial\over\partial x}\right)=\lambda L_z,\\
P=x, \ \ F=\lambda y.
\end{eqnarray}

The Casimir operator (\ref{Kaz1}) in this representation reads
\begin{eqnarray}
K=p^2+{1\over\lambda^2}F^2=x^2+y^2.
\end{eqnarray}
Returning to nonlinear representation we find $K=1/\lambda^2$.

As result of (\ref{XPF}) the linear algebra (\ref{lin-a1}) with $\beta=-\lambda^2$
is in fact the algebra of angular momentum $L_z$ and position operators $x$, $y$.

Let us consider in frame of this algebra eigenvalue problem for $X$
that $X$ is proportional to $z$-component of the angular momentum.
In spherical coordinates the operators read
\begin{eqnarray}
x={1\over\lambda}\sin\phi,\ \ y={1\over\lambda}\cos\phi, \ \ L_z=-i{\partial\over\partial\phi}.
\end{eqnarray}
The eigenvalue equation for $L_z$
\begin{eqnarray}
-i{\partial\over\partial\phi}\psi=m\psi
\end{eqnarray}
give the solution
\begin{eqnarray}
\psi=Ce^{im\phi}.
\end{eqnarray}
The boundary conditions (\ref{bcn0}) and (\ref{bcn0m}) now read
\begin{eqnarray}\label{B1}
\psi(-\pi/2)=\psi(\pi/2), \\
\psi(-\pi/2)=-\psi(\pi/2)\label{B2}.
\end{eqnarray}
For (\ref{B1}) we find $m=2n$, where $n=0,\pm1,\pm2,...$ and thus eigenvalues
for $X$ is $l_n=\lambda2n$, that is in agreement with (\ref{eigval1p}).
For (\ref{B2}) we find $m=2n+1$, where $n=0,\pm1,\pm2,...$ and thus eigenvalues
for $X$ in this case is $l_n=\lambda(2n+1)$, that is in agreement with (\ref{eigval1m}).

\section{Algebra of total angular momenta and nonlinear algebra}
In this section we consider nonlinear deformed algebra of the form
\begin{eqnarray}\label{algXP}
[X,P]=i\sqrt{1-(\lambda_1^2X^2+\lambda_2^2P^2)}.
\end{eqnarray}
It is interesting that this nonlinear algebra is related with the Lie algebra for angular momentum
\begin{eqnarray}\label{angul}
[J_x,J_y]=iJ_z,\ \ [J_z,J_x]=iJ_y, \ \ [J_y,J_z]=iJ_x.
\end{eqnarray}
Squared total angular momentum $J^2$ commute with each component of the total angular momentum and is the Casimir operator.
Let us consider the subspace with fixed eigenvalue of squared total angular momentum.
Then
\begin{eqnarray}
J^2=J_x^2+J_y^2+J_z^2=j(j+1),
\end{eqnarray}
where $j=0,1,2,...$ or $j=1/2, 3/2, 5/2,...$.
We find
\begin{eqnarray}\label{Jz}
J_z=\pm\sqrt{j(j+1)-(J_x^2+J_y^2)}.
\end{eqnarray}
Choosing $"+"$ and substituting it into the first equation in (\ref{angul}) we obtain
\begin{eqnarray}
[J_x,J_y]=i\sqrt{j(j+1)-(J_x^2+J_y^2)}.
\end{eqnarray}
Note that with $"+"$ we are restricted to the subspace spanned by eigenstates of $J_z$ with positive eigenvalues.

As we see, this algebra is very similar to (\ref{algXP}). Let us introduce operators of position and momentum as follows
\begin{eqnarray}\label{XPang}
X=\lambda_2 J_x, \ \ P=\lambda_1 J_y.
\end{eqnarray}
Then
\begin{eqnarray}
[X,P]=i\lambda_1\lambda_2\sqrt{j(j+1)-\left({1\over\lambda_2^2}X^2+{1\over\lambda_1^2}P^2\right)}=\\
=i\sqrt{\lambda_1^2\lambda_2^2j(j+1)-\left({\lambda_1^2}X^2+{\lambda_2^2}P^2\right)}.
\end{eqnarray}
Choosing
\begin{eqnarray}\label{l1l2}
\lambda_1^2\lambda_2^2j(j+1)=1
\end{eqnarray}
we obtain deformed algebra (\ref{algXP}). So, nonlinear deformed algebra (\ref{algXP}) is related with
the Lie algebra for total angular momentum (\ref{angul}).

Finally, let us introduce new operator
\begin{eqnarray}\label{FJ}
F=\lambda_1\lambda_2 J_z=\sqrt{1-(\lambda_1^2X^2+\lambda_2^2P^2)}.
\end{eqnarray}
Then according to (\ref{XPang}) and (\ref{FJ}) the algebra of operators $X, P, F$ reads
\begin{eqnarray}\label{LinJ}
[X.P]=iF, \ \ [X,F]=-i\lambda_2^2P, \ \ [P,F]=i\lambda_1^2X.
\end{eqnarray}
Thus, the nonlinear algebra (\ref{algXP}) with the help of operator (\ref{FJ}) can be extended to linear one (\ref{LinJ}).
It is worth to stress that parameters $\lambda_1$ and $\lambda_2$ are not independent but are related by
(\ref{l1l2}).

Here it is worth to stress that in the limit $\lambda_1\to 0$ the linear algebra (\ref{LinJ}), which
is related to the algebra of angular momentum,
corresponds to (\ref{lin-a1}), which
is related to the algebra of transformations of the Euclidian space. This limit corresponds to the contraction procedure
described in \cite{Gilmore} (Chapter 13).
Respectively, in the limit $\lambda_1\to 0$ deformed algebra (\ref{algXP}) corresponds to (\ref{a1}) with the function of deformation (\ref{solf11}). Therefore, the contraction procedure relating two linear algebras relates also corresponding
nonlinear deformed algebras.

Using representation (\ref{XPang}) one can find that eigenvalues of position operator are $\lambda_2 m$, where $ -j\le m\le j $
and minimal length is zero. Similarly, the eigenvalues for momentum operator are $\lambda_1 m$ and thus minimal momentum is zero.

In conclusion it worth stressing that proposed here construction of nonlinear algebra from linear one
with the help of the Casimir invariant can be applied to an arbitrary linear algebra.

\subsection{Harmonic oscillator}
We consider the eigenvalue problem for harmonic oscillator with Hamiltonian
\begin{eqnarray}\label{Hosc}
H={1\over 2}(P^2+X^2)
\end{eqnarray}
in space described by deformed algebra (\ref{algXP}) where $\lambda_1=\lambda_2=\lambda$
\begin{eqnarray}\label{algXP1}
[X,P]=i\sqrt{1-\lambda^2(X^2+P^2)}.
\end{eqnarray}
This deformed algebra is related with the Lie algebra for total angular momentum when (\ref{l1l2})
is satisfied. It gives the relation
\begin{eqnarray}\label{lj}
\lambda^4={1\over j(j+1)}.
\end{eqnarray}

In order to find energy spectrum of (\ref{Hosc}) we use the relation of nonlinear algebra (\ref{algXP1}) with the
Lie algebra for total angular momentum (\ref{angul}). Using (\ref{XPang}) we rewrite the Hamiltonian as follows
\begin{eqnarray}\label{HJ}
H={\lambda^2\over 2}(J_x^2+J_y^2)={\lambda^2\over 2}(J^2 - J_z^2),
\end{eqnarray}
where $J^2$ and $J_z$ commute, the
eigenvalue of $J^2$ is $j(j+1)$ and the eigenvalue of $J_z$ is $m$, $ -j\le m\le j $. Note that on the subspace
where deformed algebra is related with Lie one the quantum number $j$ is fixed and is related
to the deformed parameter (\ref{lj}), $m$ is positive that corresponds to choosing $"+"$ in (\ref{Jz}).

Thus eigenvalues of (\ref{HJ}) read
\begin{eqnarray}
E_m={1\over 2\sqrt{j(j+1)}}(j(j+1)-m^2),
\end{eqnarray}
where for integer $j$ the quantum number $m=0,1,2,...,j$, for half integer $m=1/2, 3/2,...,j$.

Note that in this notation the maximal quantum number $m=j$ corresponds to the ground state energy.
It is convenient to rewrite $m=j-n$ where $n=0$ corresponds to the ground state energy.
Then
\begin{eqnarray}
E_n={1\over 2\sqrt{j(j+1)}}(j(j+1)-(j-n)^2),
\end{eqnarray}
where for integer $j$ the quantum number $n=0,1,2,...,j$, for half integer $n=0, 1,...,j-1/2$.

In the limit $j\to \infty$ the deformed parameter $\lambda \to 0$ and for the energy spectrum we obtain
\begin{eqnarray}
E_n=n+{1\over 2}.
\end{eqnarray}
It reproduces the energy spectrum of a non-deformed harmonic oscillator as it must be
when the parameter of deformation tends to zero.

\section{Conclusions}

In this paper we establish the relation between nonlinear algebras and linear ones.
Namely, we find that deformed nonlinear algebra (\ref{a1}) for two operators with function of deformation (\ref{solf11})
can be transformed to linear algebra (\ref{lin-a1}) with three operators. It is interesting to note that this
linear algebra is equivalent to the Lie algebra of angular momentum $L_z$ and coordinates $x$, $y$.
The next interesting fact revealed for algebra (\ref{a1}) with function of deformation (\ref{solf12}) is
that here we have two possible scenarios concerning the existence of the minimal length.
When zero boundary condition (\ref{bc0}) is imposed then there exists nonzero minimal length
and there are no solutions of the eigenvalue equation for position operator.
For nonzero boundary conditions (\ref{bcn0}) and (\ref{bcn0m}) there exist solutions of eigenvalue equation for the position operator and
the eigenvalues are discrete.
The minimal length which is defined as a nonzero minimal uncertainty in position in this case obviously is zero.

We also show that starting from linear algebra it is possible to find corresponding nonlinear one. Namely, starting from the Lie algebra
for total angular momentum we construct corresponding nonlinear algebra for two operators which can be associated with
position and momentum operators.
The relation between linear and nonlinear algebras is not only interesting on its own right but is important from the practical point of view.
This relation gives a possibility to simplify the eigenvalue problem for corresponding operators.
It is demonstrated in section 3.1 on the example of harmonic oscillator with deformed algebra. Using the relation
of this algebra with the algebra of total angular momentum we easily find the energy spectrum for this oscillator.
\section*{Acknowledgment}
VMT thanks for warm hospitality the University of Zielona G\'ora where the main part of this paper was done.

\end{document}